\documentstyle[proceedings,namedreferences]{crckapb}
\begin{opening}
\title{Looking into Pulsar Magnetospheres}
\author{Maxim Lyutikov}
\institute{Canadian Institute for Theoretical Astrophysics\\
60 St. George street, Toronto, Canada}
\end{opening}

\newcommand{\be}{\begin{equation}}
\newcommand{\ee}{\end{equation}}
\newcommand{\ba}{\begin{eqnarray}}
\newcommand{\ea}{\end{eqnarray}}
\newcommand{\apj}{ApJ}

\begin{document}

\begin{abstract}
Diffractive and refractive  magnetospheric scintillations may allow a 
direct testing of the plasma inside the light cylinder.
Unusual electrodynamics of  the strongly magnetized electron-positron plasma
allow  separation of the magnetospheric and interstellar scattering.
 The most distinctive feature of the magnetospheric
scintillations is their independence on frequency.
Diffractive scattering due to small scale inhomogeneities
 produce a scattering angle that may be as large as $0.1$ radians, 
and a typical  decorrelation time of $10^{-8}$
 seconds.
 Refractive  scattering  due to large  scale inhomogeneities
is also possible, with a typical angle of $10^{-3}$ radians and a correlation time of the
order of $10^{-4}$ seconds.  Some of the  magnetospheric
 propagation effects may have already been
observed.
\end{abstract}

A number of observational results may possibly be attributed to scattering
processes inside pulsar magnetospheres. The most convincing are the  results of
Sallmen et al. (1999) in which  the {\it frequency independent} 
 spread and the multiplicity of the
Crab giant pulses
 with  large variations in the pulse broadening times. Other  observations
that can be possibly attributed to magnetospheric propagating effects
include  an unusual scaling of broadening times for  giant pulses 
(Hankins \& Moffett 1998), "ghosts" images of the Crab pulse and interpulse
(Smith \& Lyne 1999, Backer 1999), enhanced intensity fluctuations at very high
frequencies (Kramer et al. 1997),
enhanced scattering of nearby pulsars (Gupta et al. 1994,
Rickett et al. 1999),
 comparatively large size of the Vela pulsar's radio emission region
(Gwinn et al. 1997), absence of mictrostructure with 
very short time scales (Popov 1999), 
frequency independent scintillations with bandwidth 4~MHz (Gwinn et al. 1999).

Strong magnetic fields present in pulsar magnetospheres and the unusual electrodynamics of the
one-dimensional electron-positron plasma both
change the
familiar effects of scattering and refraction in plasma.  The unusual
features of  scattering   in such plasma may allow separation from the
interstellar scattering and will serve as a tool to probe the structure
of the magnetosphere itself.
Typically,  the frequency of the observed radio waves is much
less that the cyclotron frequency $\omega \ll \omega_B$. 
For such frequencies
the refractive index of 
strongly magnetized pair  plasma
for escaping electromagnetic modes is approximately
\begin{equation}
\delta \equiv  n^2-1\approx { \gamma_p \omega_p^2 \over \omega_B^2}  =
 3  \times \, 10^{-4} \left( { P \over 0.1 {\rm \, sec}}\right)^2 \,
\label{delta}
\end{equation}
Thus, the large Lorentz factor of the moving plasma effectively  enhances the wave-plasma interaction on the o
pen field lines.
Parameter $\delta$ is the key to the scattering and diffraction effects in the
pulsar magnetosphere. An important  fact is that $\delta$ does  not strongly
depend on our assumptions about the density and the streaming Lorentz factors of the
plasma.

 Parameter $\delta $, which determines refractive
properties of the medium,  is negligible deep
inside the pulsar magnetosphere, but
increases with the distance from the neutron star
as $\propto r^3$. Thus, the strongest nonresonant wave-plasma interactions
occur in the outer regions of pulsar  magnetospheres (near the light cylinder).
This  allows for  a  considerable simplification  when considering scattering
and diffraction effects since one can  adopt a "thin screen" approximation.
We assume that emission is generated deep in the pulsar magnetosphere and then
scattered in a thin screen
  located
near the light cylinder with a typical thickness $D \approx 0.1 R_{LC}$.

Two types  of inhomogeneities that should be present inside the pulsar magnetosphere:
small scale inhomogeneities with a  typical sizes comparable to
tens of skin depth
 $\sim c/\omega_p$ which arise due to the plasma turbulence
 and large scale inhomogeneities with a  typical sizes comparable to
light cylinder radius which arise due to temporal and spatial modulation of the
 outflowing pair plasma.
The two types  of inhomogeneities 
will produce qualitatively different effects:  small scale inhomogeneities will produce
diffractive scattering, while large scale inhomogeneities 
will produce  refractive  scattering. 
As usual, the scale that defines "small" and "large" is the Fresnel scale
which in this case is equal to
$r_f = \sqrt{ R_{LC} \lambda} \sim 10^5 $ cm.

Assuming that scattering inside the pulsar magnetosphere is strong,
the typical scattering angle can be estimated as
\begin{equation}
\theta_{\rm \, scat} \approx { \Delta \phi \over 2 \pi} {  \lambda \over  a }=
\left(  {D\over a} \right) ^{1/2} \delta \,
 \Delta   n_e.
\label{thetascat}
\end{equation}
where $D \sim 0.1 R_{LC}$ is a thickness of the scattering screen,
$a$ is a typical size  of irregularities and $\Delta   n_e$ is a relative amplitude
of density fluctuations.

A  necessary requirement to allow for
 the separation of scintillations into diffractive and
refractive branches is that the scattering  should be  strong, so that the
total phase shift of the wave is much larger than $\pi$. For a
medium with given size inhomogeneities the scattering is strong for
\be
\lambda <  \lambda {\rm \, max}  = \sqrt{ a D} \delta \approx
\left\{ \begin{array}{ll}
30 \, {\rm \, cm} & \mbox{ for $a  = a_{\rm \, min} $}\\
10^5\,  {\rm \, cm} & \mbox{ for $a  = a_{\rm \, max} $}
\end{array} \right.
\ee
This is a limitation
of the wavelength {\it from above}:  scattering is stronger for shorter wavelengths.
This is in sharp contrast to the scattering in the interstellar medium, where the
strength of the scattering increases at low frequencies.

For the typical wavelength of observations $\lambda = 30 {\rm \, cm}$,
both refractive and diffractive scattering occurs in a strong scattering regime
with the  total phase change:
\be
\Delta \phi \approx  \left\{
\begin{array}{ll}
1 & \mbox{ diffractive}\\
100 & \mbox{ refractive}
\end{array}
\right.
\ee
Diffractive scattering become weak at longer wavelengths.

Assuming that there are strong density fluctuations,  $\Delta n_e \sim 1$,
present on smallest scales $a \sim  a_{\rm \, min}$
the diffractive scattering angle becomes $\theta_{\rm \, D} \approx 0.1 {\rm rad}$.
This implies that large angle scattering is possible in the outer regions
of pulsar magnetospheres.
If this extreme scattering case was realized, the observed profiles  are  then would
be a convolution of the "initial" window function
(determined by the emission conditions at lower radii)
with  diffractive scattering.
Since this is not what is observed, we should collude that strong density fluctuations
 are not present at very small scales.

On the other hand, strong
refractive density fluctuations fluctuation are almost  definitely present 
inside the pulsar magnetosphere. 
Typical
 refractive scattering angle is
\begin{equation}
\theta_{\rm \, R} \approx \delta = 4  \times \, 10^{-3}.
\label{thetaref}
\end{equation}
Refractive effects will induce "jitter" in the arrival times of the pulses and
a temporal
correlation in the intensities with a
typical scale $\tau_R = \theta_{\rm \, R} P = 4  \times \, 10^{-4} {\rm \,  sec}$.
Both of these effects will be independent of frequency,
and increasing with the period of the pulsar.

Scattering may also affect polarization structure of emission. 
If two regions in the magnetosphere emit different polarization, scattering
may mix  the radiation decreasing the degree of polarization. 
The amount of mixing will be different in the weak scattering regime
 at frequencies $\leq $ 1 GHz and strong  scattering regime at $\geq $ 1 GHz. 
This effect may be responsible for the
  high frequency depolarization observed in some pulsars
(I thank Barney Rickett
for pointing out this possibility). 
Also note, that the real change of  the polarization vector of the
radiation is impossible in the outer regions of the magnetosphere, beyond the
limiting polarization surface.

Another possible effect of propagation inside the pulsar magnetosphere
include a reflection from the boundary between the closed and open field
lines. Plasma density on the closed field lines may approach a value at which the
thermal energy density of plasma reaches magnetic field energy density. 
For mildly relativistic plasma with thermal velocity of the order of velocity of light
the  maximum plasma density ($ n \sim B^2 /( 8 \pi m c^2)$)

Experiments to detect effects
 of wave propagation inside the pulsar magnetosphere should use high frequencies
and concentrate on nearby
pulsars with low dispersion measure.
Possible  experiments  will include a search for nondispersive 
effects such as a
time delay (as large as tens of microseconds) in the pulse arrivals,
a  diffractive   decorrelation bandwidth of the order of $10$ MHz,
 and microstructure periodicities (of the order of tens of microseconds)  due to refractive scattering.
Nearby strong pulsar,  
like PSR 0950, are best   candidate for searches for magnetospheric
effects.

The predicted characteristics of the scattering  inside
the pulsar magnetosphere are
\begin{tabbing}
diffractive scattering angle \hskip .3 truein \=  $  10^{-1 } $ \\
diffractive scattering  time \> $10^{-4} {\rm \, sec}$ \\
diffractive decorrelation time \> $10^{-8  }  {\rm \, sec}$ \\
refractive  scattering angle  \>   $ 10^{-3  } $ \\
refractive  decorrelation time \> $10^{-4 } {\rm \, sec} $ \\
arrival time variations \> $10^{-4 } {\rm \, sec}$ \\
\end {tabbing}

\end{document}